\documentclass[oldversion]{aa}
\usepackage{graphicx}
\usepackage{graphpap}
\usepackage{epsf}
\usepackage{amssymb}
\usepackage{color}

\usepackage{natbib}

\def\FIGw #1 #2 [#3] #4\par{%
  \begin{figure*}\begin{center}%
    \includegraphics*[#3]{#2}%
    \\
    \caption{#4}%
    \label{#1}%
  \end{center}\end{figure*}%
}

\def\FIG #1 #2 [#3] #4\par{%
  \begin{figure}\begin{center}%
    \includegraphics*[#3]{#2}%
    \\
    \caption{#4}%
    \label{#1}%
  \end{center}\end{figure}%
}
\def\FIGG #1 #2 #3 [#4] #5\par{%
  \begin{figure}
               \includegraphics*[#4]{#2}
               \includegraphics*[#4]{#3}
    \caption{#5}%
    \label{#1}%
  \end{figure}%
}
\def\FIGth #1 #2 #3 #4 [#5] #6\par{%
  \begin{figure}[ht]\begin{center}%
        \includegraphics[#5]{#2}
        \includegraphics[#5]{#3}
        \includegraphics[#5]{#4}
        \caption{#6}
        \label{#1}
   \end{center}\end{figure}
}
\def\FIGfo #1 #2 #3 #4 #5 [#6] #7\par{%
  \begin{figure}[ht]\begin{center}%
        \includegraphics[#6]{#2}
        \includegraphics[#6]{#3} \\
        \includegraphics[#6]{#4}
        \includegraphics[#6]{#5}
        \caption{#7}
        \label{#1}
   \end{center}\end{figure}
}
\def\FIGsi #1 #2 #3 #4 #5 #6 #7 [#8] #9\par{%
  \begin{figure}[ht]\begin{center}%
        \includegraphics*[#8]{#2}
        \includegraphics*[#8]{#3}\\
        \includegraphics*[#8]{#4}
        \includegraphics*[#8]{#5}\\
        \includegraphics*[#8]{#6}
        \includegraphics*[#8]{#7}
        \caption{#9}
        \label{#1}
    \end{center}\end{figure}
}

\def\rfig#1{Fig.~\ref{#1}}

\newcommand{\mt}[1]{\mathrm{#1}}

\def\lvm{\leavevmode\hbox to\parindent{\hfill}}

\def\BE{\begin{equation}}
\def\EE{\end{equation}}
\def\BA{\begin{array}}
\def\EA{\end{array}}
\def\BAN{\begin{eqnarray}}
\def\EAN{\end{eqnarray}}

\def\fun#1#2{\lower3.6pt\vbox{\baselineskip0pt\lineskip.9pt
\ialign{$\mathsurround=0pt#1\hfil##\hfil$\crcr#2\crcr\sim\crcr}}}

\def\ltsima{$\; \buildrel < \over \sim \;$}
\def\ltsim{\lower.5ex\hbox{\ltsima}}
\def\gtsima{$\; \buildrel > \over \sim \;$}
\def\gtsim{\lower.5ex\hbox{\gtsima}}

\usepackage{txfonts}
\usepackage{natbib}
\bibpunct{(}{)}{,}{a}{}{,}

\begin{document}

\title{The kinematics and chemical stratification of the Type Ia supernova remnant 0519-69.0}
\subtitle{An XMM-Newton and Chandra study}
\titlerunning{Kinematics and  chemical stratification of SNR 0519-69.0}

\author{D.~Kosenko\inst{1,2} \and E.A.~Helder\inst{1} \and J.~Vink\inst{1}}
   \institute{
Astronomical Institute Utrecht, Utrecht University, P.O. Box 80000, 3508TA Utrecht, The Netherlands 
         \and
Sternberg Astronomical Institute, 119992, Universitetski pr., 13, Moscow, Russia \\
    \email{D.Kosenko@uu.nl}
             }

\authorrunning{Kosenko et al.}
\date{Received 18 December 2009 / Accepted 14 March 2010}

\abstract{We present a detailed analysis of the XMM-Newton and Chandra X-ray
data of the young Type Ia supernova remnant SNR~0519-69.0, which is situated in
the  Large Magellanic Cloud. We used data from both the
Chandra ACIS and XMM-Newton EPIC MOS instruments, and high resolution
X-ray spectra obtained with the XMM-Newton reflection grating spectrometer
(RGS).

Our analysis of the spatial distribution of X-ray line emission using
the Chandra data shows that there is a radial stratification of oxygen,
intermediate mass elements (IME) and iron, with the emission
from more massive elements peaking more toward the center.
Using a deprojection technique we measure a forward shock radius of
$4.0\pm0.3$~pc and a reverse shock radius of $2.7\pm0.4$~pc.

We took the observed stratification of the shocked
ejecta into account in the modeling of
the X-ray spectra, for which we used multi-component
non-equilibrium ionization models, with the components corresponding
to layers dominated by one or two elements. An additional component was
added in order to represent the shocked interstellar medium, which mostly
contributed to the continuum emission.
This multicomponent model fits the data adequately, and was also
employed to characterize the spectra of distinct regions extracted from
the Chandra data. From our spectral analysis we find that the approximate fractional
masses of shocked ejecta for the most abundant elements are:
$M_\mt{O} \approx 32$\%, $M_\mt{Si/S} \approx 7\%/5\%$,
$M_\mt{Ar+Ca} \approx 1$\% and $M_\mt{Fe} \approx 55$\%.
From the continuum component we derive a circumstellar density of
$n_{\rm H}= 2.4\pm 0.2$~cm$^{-3}$.
This density, together with the measurements of the forward and reverse
shock radii suggest an age of 0519-69.0 of $450\pm 200$~yr,
somewhat lower than, but consistent with the age estimate based
on the extent of the light echo ($600\pm200$~yr).

Finally, from the high resolution RGS spectra we measured a Doppler
broadening of $\sigma = 1873\pm 50$~km\,s$^{-1}$, from which
we derive a forward shock velocity of $v_{FS} =2770 \pm 500$~km\,s$^{-1}$.
We discuss our results in the context of single degenerate explosion models,
using semi-analytical and numerical modeling, 
and compare the characteristics of 0519-69.0 with those of other Type Ia supernova remnants.
}


\maketitle

\section{Introduction} 
Thermonuclear (type Ia) supernova explosions have drawn a lot of attention
over the last decade, as they provide a powerful tool to measure
cosmological distances \citep{perlmutter99,riess98}. The reason is that they are
bright enough to be observed over large distances, and that,
compared to core collapse supernovae, 
their peak luminosity shows relatively little variation, which can be
further reduced by applying the empirical peak luminosity --- decline rate 
correlation  \citep{phillips99}. 
Nevertheless, some evolutionary systematic effects \citep{panagia05} may 
take place, which may affect the measured
values of cosmological parameters.
Thus, it is important to understand the mechanism and physics that governs 
thermonuclear explosions. 

In addition to the extensive studies of the supernovae themselves,
through their  light curves 
\citep[e.g][]{woosley07, blinn06} and spectra \citep[e.g.][]{branch}, 
one can also turn the attention to the SN type Ia remnants. 
A typical young supernova remnant (SNR) is a bright X-ray source, 
due to the high temperatures of the plasma, heated by the forward and reverse 
shocks. The reverse shock, propagating inwards into the supernova ejecta, 
efficiently heats the metal-rich matter. The hot plasma produces X-ray spectra,
abound with prominent emission lines. These spectra carry the imprints of the 
chemical composition of the plasma, and, therefore,  the 
distribution of the elements in the 
supernova ejecta, which are determined by the explosion properties. 
Thus, a detailed analysis of the 
X-ray data of supernova type Ia remnants provides a powerful complementary 
tool in the studies of the nature of a thermonuclear supernova progenitor 
\citep[e.g.][]{sorokina,badenes03}.

The current generation of X-ray observatories, such as XMM-Newton and Chandra, 
provide the capability to obtain simultaneously spectral and imaging data of 
extended objects such as SNRs. Young SNRs in the Large Magellanic Cloud (LMC) 
are especially suitable targets for these telescopes, as the relative proximity 
\citep[48 kpc, e.g.][]{lmc1, lmc2} of the LMC yields a sufficiently high 
signal-to-noise level of the X-ray data as observed by the CCDs and 
grating spectrometers on board Chandra and XMM-Newton. The SNRs are still 
large enough (typically  $\sim 1'$) to perform morphological studies with 
Chandra and small enough to be excellent targets for the Reflection Grating 
Spectrometer of XMM-Newton.  An additional advantage is the 
relatively low interstellar absorption column towards LMC  SNRs
($N_{\rm H}\approx 10^{21}$~cm$^{-2}$), as compared to most Galactic SNRs.

\begin{figure}
\begin{center}%
        \includegraphics[width=0.7\hsize]{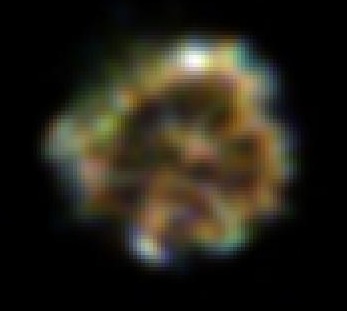}
        \caption{Chandra smoothed RGB image of SNR~0519-69.0. Red --- 0.5 - 1.0 keV, green --- 1.0 - 2.0 keV, blue --- 2.0 - 4.0 keV. North is up, east is left. } 
	\label{chandra_rgb}
\end{center}\end{figure}

In this paper we present an X-ray study of the LMC 
SNR~0519-69.0. 
The SNR has an irregular patchy morphology
and an angular extent in X-rays of 33\arcsec, 
corresponding to a shell radius of  4~pc. 
A three color X-ray image of SNR~0519-69.0, based on the Chandra data,
is presented in \rfig{chandra_rgb}. 

In the optical SNR~0519-69.0 has been 
investigated by \citet{tuohy82, smith91} and \citet{ghav07}. It is also one of
several LMC SNRs for which a light-echo has been identified, from
which an age of $600\pm200$ years can be deduced \citep{rest}.

In X-rays the remnant was studied by \citet{hughes95}, who analyzed the ASCA 
spectra and found that the SNR is oxygen-poor and iron-rich and must be a remnant of a 
thermonuclear supernova (SN Ia) explosion. The analysis of Chandra data 
by \citet{williams01} revealed the separation between the shocked ejecta and 
the shocked circumstellar medium (CSM).

In the present study we analyzed 
archival X-ray data of SNR~0519-69.0 from the both 
XMM-Newton and Chandra observatories.  We used the ACIS data of Chandra 
and  the imaging spectroscopy (EPIC) and the high spectral 
resolution grating (RGS) data of  XMM-Newton.

Various techniques to analyze the available data are employed. 
The high spatial resolution Chandra data provide us 
with an opportunity to study the composition of the SNR as a function
of radius. Whereas the RGS data offer high resolution spectroscopy, but
without much spatial information.

For the analysis of the SNR~0519-69.0 X-ray spectra, we employed the {\sc spex} 
\citep{spex} spectral fitting software (version 2.01.05, November 16, 2009). The package contains 
the most up-to-date atomic data and has a wide range of plasma emission models,
which is especially helpful and important in fitting complex spectra from 
objects such as SNRs.

We fit the XMM-Newton EPIC and RGS spectra with single and 
multicomponent NEI models. The high spectral resolution RGS data 
allow us to resolve details of the Fe-L line emission, and 
to measure line velocity broadening, due to the thermal and bulk motion of
the shocked supernova ejecta.

This paper is organized as follows. First  in Sect.~\ref{data}, we describe 
briefly the Chandra  and XMM-Newton data. The methods and techniques we used
 to deal with the spectral and imaging data are presented in 
Sect.~\ref{analysis}. Sect.~\ref{results} contains the principal results of 
the study, which are discussed in Sect.~\ref{discussion}. 
We conclude the paper in Sect.~\ref{conclusion}.

\section{The data overview} \label{data} 
\subsection{Chandra} 
SNR~0519-69.0 was observed only with the ACIS instrument of Chandra X-ray 
Observatory (obs ID 118) for 41.1 ks on June 21, 2000.
We analyzed the data with CIAO 4.1.1 software product and CalDB 4.1.2 calibration data.
Apart from an X-ray spectrum of the whole SNR, we also used our own  
software to extract spectra from certain regions based on emission
characteristics. The software reads the standard pipeline event lists and
selects events from it based on a ``mask'' image, i.e. an image containing
only zeroes and ones. An event is selected for spectral extraction, if
its sky coordinates corresponds with a pixel of value one in the mask image.
The event grade selection and spectral data binning scheme are 
identical to the  standard CIAO software. The advantage of this method 
is that spectra can be extracted from regions with more complex
shapes than with the region files, and can be easily constructed based
on hardness ratios or narrow band images.
These spectra were used together with the ancillary and main response 
files generated for the SNR as a whole. Given that the remnant occupies only
a small part of the ACIS-S3 CCD, using these files for the response
is justified. For the background spectrum, we extracted a spectrum
from an annulus with radii spanning from 19\arcsec to 37\arcsec, 
centered on the SNR.

\subsection{XMM-Newton} 
We used  XMM-Newton observations of SNR~0519-69.0 (obs ID~0113000501) that 
were made on September 17, 2001, and have a total exposure of 47.8 ks.

For our study  we concentrated on the EPIC MOS and RGS data.
Although the EPIC MOS instruments \citep{turner01}
have a lower sensitivity than the EPIC pn 
instrument, they have a higher spectral resolution, which is important for 
line-rich sources, such as SNRs.
The MOS1 observations were performed with thick (25.4 ks) and medium (15.6 ks) 
filters,  whereas the MOS2 data were acquired with the medium filter only. 
For the spectral analysis the spectra, response and ancillary response files 
were combined, using weights proportional to the exposure times. 
For the background, spectra were extracted from an annulus with an inner
radius of 2.1\arcmin, and an outer radius of 4.3\arcmin. This is larger than
for the Chandra background spectrum, because the XMM-Newton point spread
functions have much broader scattering wings.

The RGS is a slitless spectrometer \citep{denherder}. 
For an extended source, this implies that 
the spectrum is smeared by the image of the source itself.
For Small/Large Magellanic Cloud remnants, the smearing is modest, but 
present, and it gives rise to a change in the line spread function. For our 
analysis, we incorporated this effect into the response matrix by convolving 
the standard (point source) response matrix with the brightness profile of the 
SNR, as obtained from the Chandra observations. All the results presented here 
made use of these modified matrices. This procedure was also applied to the RGS data of the
SN~1006 remnant \citep{vink03} and SNR~0509-67.5 \citep{0509kosenko}. 

Apart from adapting the RGS response matrix, all reduction for both MOS and 
RGS data  were made with the standard XMM-Newton software package SAS 
version~7.1.0.

\section{Data analysis}\label{analysis}
\subsection{Chandra radial emissivity profiles}\label{sec_profiles}
The high spatial resolution Chandra data allow us to plot radial emissivity
profiles of the remnant's shell, with a deprojection technique that was also
employed by \citet{helder08}.  For the measurements, we first made a radial
surface brightness profile of the remnant, using the centroid
of the broadband emission (0.5-8.0 keV) as center.
Then, assuming that the SNR is
spherically symmetric, we used the
Lucy-Richardson technique \citep{lucy74, richardson72} to deproject the
surface brightness profile into an emissivity profile
\citep[following the method of][]{willingale96,helder08}.
We applied this technique to images in different energy bands,
where the emission lines from different species dominate.
For the oxygen band, we choose 0.5-0.7 keV,
for the added silicon and sulfur band, we took 1.75-3.00 keV.
Because the spectra indicated  that in the Fe-L band
($\sim 0.7-1.1$~keV) the relative contribitions from different Fe-L shell
ions varies as a function of position (Sect.~\ref{sec_masks}),
we divided the iron energy band in a low (0.7-0.9 keV) and high (0.9-1.1 keV)
energy band, which correspond roughly to line emission dominated by
Fe~XVII and Fe~XX ions respectively.

For the broad band (0.5-8.0 keV)  energy range, we calculated an inner and outer radius.
The inner radius was determined where
the emissivity (i.e. deprojected, $\epsilon$) drops below one half of the
maximum value.  We ignored emissivities at radii less than 7\arcsec,
since the deprojection
technique is not well constrained at these lower radii, due to the low
contribution to the surface brightness $\Sigma$ of shells at small radii
($\Sigma \propto \epsilon(R) R^2$).
In general, there is no sharp rise in emissivity at the outer radius,
so taking a value of one half of the maximum would underestimate the outer radius.
Hence, we took a value of 1/20 of the peak value, which,
in the deprojections, is just above the noise level of the background at large
radii. In the remainder of this paper we assume that this measured inner radius of $2.7\pm0.4$~pc, 
corresponds to the location of the reverse shock and the outer radius
to the forward shock at $4.0\pm0.3$~pc. 
We estimated the error from the standard deviation of this measurement repeated
 for 18 individual slices of $20^\circ$ each. 

We use the deprojected profiles to determine the shells in the remnant in which different species dominate (\rfig{elem_profiles}), which we will use in section \ref{results} for estimating total masses of the shocked species. Table \ref{radii} lists the ranges in which the different components dominate. Silicon and sulfur dominate the emissivity in region 2/3, however, our numerical models (see Sect.(\ref{discussion})  show that the Si/S and Ar/Ca layers coincide. For this reason, we split region 2/3 evenly in two parts. Furthermore, we chose 3.6~pc as the division between oxygen and the shocked CSM, since for this value, the volume of the shocked CSM is one fourth of the volume of the total remnant. The latter seems reasonable, if we assume a compression ratio of 4 over the shock front during the whole lifetime of the remnant.
Note that all values mentioned above contain uncertainties. For the corresponding volumes, 
we estimate errors of 50\%.
In \rfig{elem_profiles} we also show the observed
radial profile and its deprojection for the 5-7 keV hard X-ray band.
This band includes both continuum emission and Fe-K
shell line emission around 6.5 keV. Unfortunately, the signal to noise ratio
in this band is poor, but the profile suggests tentatively that the 5-7 keV
emission is associated mostly with the Fe XX line emission in the 0.9-1.1 keV 
band. With some caution, it also suggests  that there is no strong
non-thermal X-ray continuum associated with the forward shock.

\begin{table*}
\caption{The inner and outer radii for the SNR~0519-69.0 shell in different energy ranges. }
\begin{center}
\begin{tabular}{llllllll}
Radii  &  0.5-8.0 keV & CSM & O & Si/S &  Ar/Ca & Fe-high & Fe-low   \\
 & & 0 & 1 & 2 & 3 & 4 & 5 \\
\hline \\
$R_{\rm in}$ (pc) & $2.7\pm0.4$ & 3.6 & 3.48 & 3.405 & 3.33  & 3.15 & 2.7\\
$R_{\rm out}$ (pc) & $4.0\pm0.3$ & 4.0 & 3.6  &3.48 & 3.405 & 3.33 & 3.15  \\ \hline
\end{tabular}
\end{center}
\label{radii}
\end{table*}

\FIGw elem_profiles whole_circle_double [width=0.9\hsize,angle=0] Azimuthally averaged radial profiles in different energy bands. Left panel: surface brightness radial profiles. Right panel: deprojected emissivity profiles. The vertical light grey lines correspond to the spectral components, used in Sect.~\ref{multinei}.

\subsection{EPIC and ACIS spectra}\label{epicacis}
Fitting the XMM-Newton EPIC MOS and Chandra ACIS spectra 
with single-ionization timescale non-equillibrium ionization (NEI) {\sc spex} 
model provides us with the typical values for the emission measure, electron 
temperature, ionization timescale and abundances in the remnant. 
The corresponding spectra and the best-fit model 
are shown in \rfig{epic_acis_nei}. 
The best-fit parameters of the spectral model are listed in 
Table~\ref{all_nei_pars}. The best-fit abundances are plotted in 
\rfig{abunds}. For comparison, also the abundances, derived from the analysis 
of the younger supernova Type Ia remnant 0509-67.5 \citep{0509kosenko} 
are plotted. 

\FIG epic_acis_nei epic_acis_nei_absmnb_sm [width=0.95\hsize,angle=0]  Combined fitting of the EPIC MOS (scaled down by a factor of 10) and ACIS (scaled up by a factor of 10) data with single ionization timescale NEI model of {\sc spex}.

A single ionization timescale NEI model is not the best approach to explain 
emission from such a complicated and layered object as a SNR, because
in the shell one expects ionization timescale and abundance gradients. For 
example, \rfig{epic_acis_nei} clearly shows that Si Ly$\alpha$ line is missing 
in the model and Fe K line flux is overestimated. This discrepancy leads 
to a very high value of the fit statistic; $\chi^2/d.o.f. \simeq 30$.

Apart from the MOS, we fitted also the EPIC pn data with an NEI model, 
that yields approximately the same values of the basic parameters of the 
spectrum.

\FIG abunds epic_acis_abunds  [width=0.9\hsize,angle=0] Best-fit abundances  of SNR~0519-69.0 and SNR~0509-67.5 \citep{0509kosenko}, derived from single-ionization timescale NEI models. The data are in solar units \citep{ag89}, normalized by Si abundances.

\begin{table*}
\caption{The best-fit NEI parameters of the XMM-Newton EPIC, RGS and Chandra ACIS data. Columns 1 and 2 list the data for the entire remnant, column 3 lists data for the inner shell of the SNR and column 4 --- for the outer shell (see \rfig{chandra_reg}).}
\begin{center}
\begin{tabular}{lllll}
\hline \\  Parameter  & RGS & EPIC MOS + ACIS & ACIS inner & ACIS outer \\
\hline \\ $n_en_H\,V\; (10^{58},\;\mt{cm}^{-3}$) &
 $1.8^{+2.5}_{-1.8}$ & $3.7_{-0.2}^{+0.2}$  & $2.63_{-0.32}^{+0.33}$  & $0.66_{-0.08}^{+6.49}$  \\
$kT_e$ (keV)                               &
$2.21^{+0.08}_{-0.07}$  & $2.82_{-0.03}^{+0.03}$ & $2.88_{-0.06}^{+0.06}$ & $1.73_{-1.07}^{+0.14}$  \\
$n_et\; (10^{10},\;\mt{s\;cm}^{-3})$     &
 $2.69^{+0.03}_{-0.03}$ & $2.27_{-0.01}^{+0.01}$ & $2.22_{-0.002}^{+0.02}$ & $3.90 \pm N/A$   \\
$n_H\; (10^{21},\;\mt{cm}^{-2}$)     & 
--- &  $2.62_{-0.06}^{+0.07}$ &  $2.74_{-0.11}^{+0.11}$ & --- \\
\hline $\chi^2/d.o.f.$ & $2.4$ & $29.4$& $10.4$ & $2.5$ \\
\hline
\end{tabular}
\end{center}
\label{all_nei_pars}
\end{table*}

\subsection{XMM-Newton RGS spectra}\label{rgsdata}
The RGS spectra of SNR~0519-69.0 were also fitted with {\sc spex} 
NEI models. The fit range was limited to the 0.5-1.1 keV range, which has
the more prominent line emission.
The best-fit parameters of the single NEI model are listed in 
Table~\ref{all_nei_pars} (first column). This single-ionization timescale 
model was not able to reproduce the fluxes of O VII and Fe XXI ions: the 
modeled lines are weaker than the data indicate (top panel of \rfig{rgs}). 
For this  reason, and inspired by the spatial layering, as indicated by the Chandra data
(Sect.~\ref{sec_profiles}), 
we fitted the RGS spectra in the range of 0.5-1.1 keV
with a three-component NEI model: one 
component is for pure oxygen, the second ---  for the low-ionized iron 
(Fe~XVII -- Fe~XVIII), and the third --- for the high-ionized iron (Fe~XIX -- 
Fe~XXI). The corresponding spectra are shown in the bottom panel of \rfig{rgs}.
The corresponding best-fit parameters are listed in the Table~\ref{rgs3_pars}. 
As can be seen in Fig.~\ref{rgs}, this model gives a much better fit to
the data than the single NEI model. Note, that no Ne emission is needed to obtain a good fit 
(Ne~IX has a prominent line at  0.92 keV and Ne~X at 1.02 keV).

\FIGG rgs rgs_nei_narrow_sm rgs_3nei_narrow_sm  [width=0.95\hsize,angle=0]  Top panel: the four RGS spectra RGS1, RGS2 of SNR~0519-69.0. Shown are  both first order (upper frame) and second order (lower  frame) spectra and best fit single NEI model including a line velocity broadening model. Bottom panel: The same as the top panel, but the spectral model consists of three NEI components. Black crosses are the data, solid lines are the models.

An important advantage of the high spectral resolution RGS spectra is that it 
enables us to measure the line broadening. 
The measured best-fit value of the line velocity broadening is 
$\sigma_v = 1680\pm50$ km\,s$^{-1}$ for the single NEI model and
$\sigma_v = 1873\pm50$ km\,s$^{-1}$\ for the multi-component model. We
adopt the last value as the most reliable value for the Doppler broadening,
since the model fits the data much better (Fig.~\ref{rgs}).

\begin{table}
\caption{The best-fit parameters of the RGS spectral fitting with the three-component NEI model. $\chi^2/d.o.f. = 1.8$ }
\begin{center}
\begin{tabular}{llll}
Parameter  & O & Fe-low & Fe-high  \\
\hline \\ $n_en_H\,V\; (10^{58},\;\mt{cm}^{-3}$) & $0.50^{+0.08}_{-0.06}$ & $5.6^{+0.2}_{-0.3}$ & $4.8^{+1.2}_{-0.7}$   \\
$kT_e$ (keV)                                       & $0.84^{+0.27}_{-0.18}$ & $1.22^{+0.05}_{-0.33}$ & $2.51^{+6.54}_{-0.99}$  \\
$n_et\; (10^{10},\;\mt{s\;cm}^{-3}$)     & $2.3^{+0.8}_{-0.5}$ & $3.0^{+2.3}_{-0.2}$ & $5.2^{+3.4}_{-1.4}$ \\ 
$\sigma_\mt{RGS} (\mt{km\;s}^{-1})$           & \multicolumn{3}{c}{$1873\pm50$} \\
\hline
\end{tabular}
\end{center}
\label{rgs3_pars}
\end{table}

\subsection{EPIC MOS and RGS spectra combined, multicomponent approach}\label{multinei}
A next step in our study was to combine both the EPIC MOS and RGS data and
try to model it with a multi-component NEI model, in which each component
roughly corresponds to the layering observed in the radial profiles 
(Sect.~\ref{sec_profiles}).
In addition to the three components used for fitting the RGS spectra
(Sect.~\ref{rgsdata}),
we added three additional components, which contribute mostly to
the emission outside the RGS spectral range. In total this amounts to six
components:
oxygen (O, or component 1), 
silicon/sulfur (Si/S, or component 2),
argon and calcium (Ar/Ca, or component 3),
high ionized iron (Fe-high, or component 4), 
low ionized iron  (Fe-low, or component 5), and a continuum dominated model
(component 0), 
for which we used an NEI model with the abundances fixed to those of the LMC
\citep{lmcrussell92}. In addition the model included an interstellar
absorption component.

The corresponding principal abundance 
parameters in components 1--5 were set to $10^7$ times the solar
value for the specific elements considered, whereas the abundances of other
elements were set to zero. Thereby we secure that the absolute abundances of the 
corresponding elements are at least two orders of magnitude higher than 
those of the hydrogen, i.e. for all practical purposes these
components correspond to pure metal plasma's. 
The corresponding spectra and the best-fit multicomponent model are presented 
in \rfig{mosrgs_mul}. The best-fit parameters (emission measure, electron 
temperature, ionization timescale) are listed in Table~\ref{multi_pars}. 
The best-fit value for the hydrogen column density is $1.7^{0.3}_{-0.1}\times10^{21}\,\mt{cm}^{-2}$.

The multi-component NEI model gives an adequate fit to the data. 
In contrast to the three NEI component model used to fit only the RGS data,
some of the oxygen line emission in this multi-component
model attributes part of the corresponding line emission to the shocked CSM 
component, in particular the O VIII line emission.
One peculiarity of the multi-component model is the high $n_{\rm e}t$ value for Ar/Ca. This is not
an artefact of the model chosen. Isolating that part of the spectrum 
that is dominated by Ar/Ca line emission and searching for other
combinations of $kT_{\rm e}$ and $n_{\rm e}t$ resulted in similar best fit
values. In general, $n_{\rm e}t$ correlates inversely with $kT_{\rm e}$. This
means that if the  $n_{\rm e}t$ of the plasma is in reality lower than our best
fit value, an unrealistically high electron temperature is required.

\FIG mosrgs_mul mos_rgs_multi1 [width=0.99\hsize,angle=0]  The EPIC MOS1 and MOS2  combined (upper frame) and  the RGS1 and RGS2  first order (lower frame) spectra with the best-fit model of the six NEI components and interstellar absorption. Black crosses are the data, the solid lines show the model. The contributions from the different components are marked with the numbers listed in Table~\ref{multi_pars}. The dotted lines correspond to the CSM component (0).

\begin{table*}
\caption{Parameters of the components in the multi NEI approach. Errors are $1\sigma$ rms,  $\Delta \chi^2 = 2$, $\chi^2/d.o.f = 7.8$. Emission measure $EM_X = n_en_XV$.}
\begin{center}
\begin{tabular}{lllllllll}
Component number  & 1 &\multicolumn{2}{c}{2} & \multicolumn{2}{c}{3} & 4 & 5 & 0 \\ 
\hline \\ 
Layer  &  O & Si & S & Ar & Ca & Fe-high & Fe-low & CSM \\
\hline \\ 
$EM_X\;(10^{54}\;\mt{cm}^{-3})$ &  
	$8.87^{+3.39}_{-0.45}$ & 
	$2.47^{+0.31}_{-0.14}$ & $1.56^{+0.20}_{-0.09}$ & 
	$0.78^{+0.63}_{-0.38}$ & $0.60^{+0.49}_{-0.29}$ & 
	$2.52^{+0.32}_{-0.35}$ & $2.15^{+0.17}_{-0.36}$ &
	$2.36^{+0.50}_{-0.11}\times10^{5}$  \\
$kT_e$ (keV)  &  
	$0.72^{+0.70}_{-0.24}$ & 
	$7.00^{+3.00}_{-2.26}$ & $7.00^{+3.00}_{-2.26}$ & 
	$2.67^{+10.00}_{-1.17}$ & $2.67^{+10.00}_{-1.17}$ & 
	$2.73^{+0.07}_{-0.27}$ & $1.26^{+0.44}_{-0.26}$ &
	$0.64^{+0.02}_{-0.06}$  \\
$n_et\,(10^{10}\,\mt{s\,cm}^{_3})$  &  
	$1.52^{+3.66}_{-0.92}$ & 
	$3.63^{+0.27}_{-0.18}$ & $3.63^{+0.27}_{-0.18}$ & 
	$27.2^{+N/A}_{-18.8}$  & $27.2^{+N/A}_{-18.8}$ & 
	$3.78^{+0.34}_{-0.26}$ & $2.55^{+0.56}_{-0.39}$ &
	$80.0_{-13.4}^{N/A}$  \\
$n_e/n_X$  & 6.6 & 21.2 & 33.7 & 30.5 & 39.6 & 19.6 & 17.5 & 1.2 \\
\hline
\end{tabular}
\end{center}
\label{multi_pars}
\end{table*}

\subsection{Spatially resolved spectroscopy}
\subsubsection{Inner and outer rings of the remnant}
In the analysis of the Chandra data we extracted ACIS spectra from two shells: 
the outer one which is presumably the shocked CSM of 
the remnant and the inner one which is the region where the emission is 
produced by the shocked ejecta material. These regions are outlined with green 
circles in \rfig{chandra_reg}. Each of the spectra was analyzed in the usual 
way and fitted with single NEI models. The corresponding best-fit parameters 
are listed in Table~\ref{all_nei_pars}. 

\subsubsection{ACIS oxygen and iron masks}
\label{sec_masks}
\begin{figure}
\begin{center}%
        \includegraphics[width=0.463\hsize]{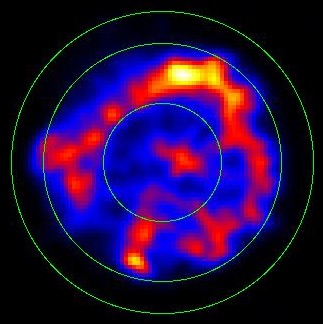}
        \includegraphics[width=0.50\hsize]{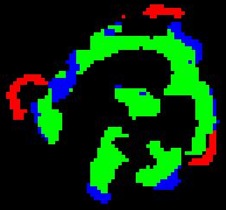}
        \caption{Left panel: Chandra image of SNR~0519-69.0. Outer and inner regions that were considered are outlined with green lines. Right panel: the masks of the Chandra image of SNR~0519-69.0: red --- oxygen ($0.5-0.7$ keV), green --- Fe-low ($0.7-0.9$ keV), blue --- Fe-high ($0.9-1.1$ keV)} 
	\label{chandra_reg}
\end{center}\end{figure}

The low energy band (0.5-1.1 keV) of the ACIS  spectra was treated more 
thoroughly.  We extracted three images of the remnant in the energy bands 
(see the RGS spectra with the corresponding emission lines) of $0.5-0.7$ keV 
(O  only), $0.7-0.9$ keV (Fe-low) and $0.9-1.1$ keV (Fe-high). Next, three 
spatial masks
were created on the basis of these images, in order
to locate regions with the relative emission peaking in either of these
energy bands. The masks are 
presented in the \rfig{chandra_reg}. They were applied to the ACIS image of the 
remnant. Finally, we extracted spectra using these, non-overlapping, 
image masks.

The resulting spectra are poorly fitted with single-ionization timescale NEI 
models, the values of the best-fit abundances contain large errors, and 
physical parameters, such as temperature and ionization timescale are of the 
order of the typical values obtained in the analysis of the EPIC and ACIS data 
of the entire remnant. This suggests that the masks did not totally separate out
specific layers, but
rather that certain layers may dominate the emission in certain
regions, with some overlap from other layers.
Therefore, we applied the six-component NEI model with the
parameters obtained in the fitting of EPIC and RGS data to each of the masked 
spectra. We fit only emission measures (i.e. normalizations) of each component 
in order to evaluate the contributions of the different components,
each corresponding to certain elements, to the spectra of these three distinct regions. 
The spectra and the best-fit models are presented in 
\rfig{mask_sp}.

\FIG mask_sp  allmasks_sp  [width=0.99\hsize,angle=-0]  The three masked ACIS spectra. From top to bottom: oxygen (data shifted upwards for three orders), Fe-high, Fe-low (data shifted downwards for three orders).

The contributions of each component to the different spectra
are presented in the \rfig{masked_norms}, in which we plot
the emission measures of each NEI component.

\FIG masked_norms mask_absa_err [width=0.99\hsize,angle=0]  The emission measure of each of the six NEI components shows the contribution of different elements to the emission of the masked regions of the Chandra data.

\section{Results}\label{results}
The fitting of the XMM-Newton EPIC MOS and Chandra ACIS spectra with {\sc spex}
 NEI models allowed us to measure (Table \ref{all_nei_pars}) the parameters of 
the plasma, such as typical electron temperature $kT_e \simeq 3$ keV, 
ionization timescale $n_et = 2.3\times 10^{10}\;\mt{s/cm}^3$ and abundances 
(plotted on \rfig{abunds}) in the remnant. The derived best-fit abundances are 
similar to those of SNR~0509-67.5~\citep{0509kosenko}, but overall the 
0519-69.0 remnant is more metal-rich (S, Ar, Ca and in particular Fe) as it is older and more of
 the supernova ejecta matter has been swept up by the reverse shock. The 
relative amount of the light elements (such as O, Ne, Mg) is comparable for 
both SNRs.

\subsection{The circumstellar matter density}
The multicomponent NEI model provides us with the parameters of the CSM component (component 0 in Table~\ref{multi_pars}).
Using the best-fit emission measure value we can make an estimate of the density of the CSM in the 
vicinity of the remnant.  
\BE
n_\mt{CSM} = \frac{1}{2}\sqrt{\frac{EM_\mt{LMC}}{V_\mt{SNR}\,n_e/n_\mt{H}}} = 2.4\pm0.2\;\mt{cm}^{-3}
\label{ncsm}
\EE
where $EM_\mt{LMC} = 2.36^{+0.50}_{-0.11}\times10^{59}\;\mt{cm}^{3}$ and $n_e/n_\mt{H} = 1.2$ 
are taken from Table~\ref{multi_pars}. We assumed, that the volume of the shocked CSM is $1/4\,V_\mt{SNR}$, with 
$V_\mt{SNR} = 4\pi/3\,R_\mt{out}^3 = 8.0\times10^{57}\,\mt{cm}^3$ and $R_\mt{out} = 4.0$ pc (Table~\ref{radii}). 

\subsection{Structure of the ejecta}
The radial emissivity profiles of the SNR in different energy bands, drawn from the Chandra observations, 
clearly show stratification of the elements in the 
remnant. \rfig{elem_profiles} shows that the outermost layer is 
oxygen rich, the next inward layer produces most of the Si and S emission, all 
the iron emission comes from the innermost layers of the shell. The iron shell 
is split in two regions with different ionization properties. The outer layer 
of high ionized (Fe XX) and the inner one of low ionized (Fe XVII) iron reflect
the effects of time-dependent ionization processes in the shocked ejecta:
the inner layer has been shocked later by the reverse shock, and therefore has the lowest
$n_{\rm e}t$.

This stratification is also confirmed by the analysis of the masked Chandra 
images (\rfig{chandra_reg} right panel) and their spectra (\rfig{mask_sp}). 
\rfig{masked_norms} shows the contribution of different elements
to the emission from  the masked images.

\subsection{The chemical composition}\label{sec_o_si_fe}
The measurements of the inner and outer radii of the shell in different energy 
bands (Table~\ref{radii}) allow us to estimate the masses of the layers, which
have been swept up by the reverse shock. Using the best-fit emission measure 
value  $EM_\mt{X} = n_\mt{e}n_\mt{X}V_\mt{X}$ of the each model of the multicomponent fitting (Table~\ref{multi_pars}), 
we can express the mass $M_\mt{X}$ of an element ``X'' (i.e. O, Si, S, Ar, Ca, Fe) as
\begin{equation}
$$
M_\mt{X} = n_\mt{X}V_{X}m_\mt{X} = \sqrt{\frac{EM_\mt{X}V_\mt{X}}{n_\mt{e}/n_\mt{X}}} m_\mt{U} A_\mt{X}
$$
\label{sweptmass}
\end{equation}
where $n_\mt{e}/n_\mt{X}$ --- electron to ion densities ratio, $m_\mt{U}$ --- atomic mass unit, $A_\mt{X}$ --- atomic mass of the element,  $V_\mt{X}$ --- volume occupied by the element (derived from the radii of Table~\ref{radii}).

Combining the available data we obtain $M_\mt{O} = 0.36\,M_\odot,\;M_\mt{Si/S} = 0.14/0.10\,M_\odot,\;
M_\mt{Ar/Ca} = 0.08/0.07\,M_\odot,\,M_\mt{Fe} = 1.05\,M_\odot,\;M_\mt{tot} = 1.8\,M_\odot$.
Note, that the mass values for the elements are affected by the $\sim 50\%$ errors of the volume 
estimates, although the total mass is less affected. 
These masses are higher than may be expected for a single degenerate Type Ia SN model, 
for which the total mass should be $M_\mt{tot}=1.4\,M_\odot$.
The most likely reason for this discrepancy may be the errors in the volume
estimates or in the filling factor. In particular, inhomogeneities such as 
knots may give emission measures skewed toward the higher density regions.
In fact, the  supernova material may be porous and clumpy, thus the emitting volume 
should be scaled down using a filling factor
\citep[e.g. see 3D simulations of thermonuclear explosions by][]{roepke08}.
For example, if we assume volume filling factor in the supernova of $\sim0.4$, 
then we get more adequate values
$M_\mt{O} = 0.23\,M_\odot,\; M_\mt{Si/S} = 0.09/0.07\,M_\odot,\; 
M_\mt{Ar/Ca} = 0.05/0.05\,M_\odot,\;M_\mt{Fe} = 0.67\,M_\odot,\;M_\mt{tot} = 1.15\,M_\odot$.
 
The multicomponent fitting of the XMM-Newton EPIC and RGS spectra reveals 
a specific behavior of the different NEI components. Table~\ref{multi_pars} 
shows that emission measure ($EM_\mt{X} = n_en_XV_X$) of each component drops 
from oxygen to calcium, while the values of the ionization timescale parameter 
$n_et$ rise from O to Ca. This trend indicates that increasing from the lighter
 to the heavier elements, the electron density overcomes the factor of time,
 which probably does not change drastically for the shocked intermediate mass 
elements (from Si to Ca). Thus, qualitatively, the decrease of the emission 
measure and increase of the ionization timescale reflect the relative 
abundances of the species from O to Ca,  i.e. the quantity 
$n_en_XV_X/(n_et) = n_XV_X/t \propto M_X$ (total shocked mass of an element X) 
is lower for the heavier species. A similar behavior of the ionization
 timescale parameter for different species was established by \citet{lewis02} 
in their analysis of the Chandra data of SNR~N103B.

From these considerations, applying the time of $t \simeq 500$ years to all components, 
we can make rough estimates of the relative contributions of the species 
in the SNR~0519-69.0 ejecta to be as follows: 
$M_\mt{O} = 32\%,\;M_\mt{Si/S} = 7\%/5\%,M_\mt{Ar+Ca} = 1\%,\;M_\mt{Fe} = 55\%$ of the total shocked mass,
which are roughly consistent with the mass ratios derived above, using volume estimates.
The relative amount of Ar and Ca is probably underestimated, due to the high value of 
ionization timescale parameter measured in the component 3, for which we do not have a good explanation
(note the corresponding errors bars in Table~\ref{multi_pars}).

\section{Discussion}\label{discussion}
\subsection{Stratification and composition of the ejecta}\label{sec_stratification}
Using the Chandra images we built azimuthally averaged radial emissivity 
profiles of the SNR in different energy bands, which clearly demonstrate the 
layering of the elements in the supernova (\rfig{elem_profiles}). Also these 
profiles reveal the effects of time-dependent ionization, as the emission from 
low ionized iron originates mostly from the inner layers of the iron shell, 
which is distinct from the high ionized ``older'' outer iron layers. 

These effects are also visible in the Chandra spectra extracted using the
masked images (\rfig{chandra_reg}). However, apparently projection effects
play a role in that the spectra of the more inward lying layers
are contaminated by the outer layers. For example,
the spectral fitting with multi-component NEI models shows (\rfig{masked_norms})
that the oxygen-rich component (component 1) 
is equally abundant in all three (O, Fe-high, Fe-low) masks. 
Si and S (component 2) is almost absent in the O-mask, as it is
inside the oxygen-rich layer, but it does contribute to the spectra
of inner layers of the Fe-low mask. Ar and Ca  (component 3) tend 
to be in the outer layers, but the parameters of this component cannot 
be reliably measured, and have considerable errors. The O-mask spectrum
hardly contains any contribution from the iron-rich components (4,5), 
as this outer layer can be isolated without much projection effects.

Fe-K emission is present in both the regions corresponding to the 
Fe-low and Fe-high masks. This is probably due to 
projection  effects, since our spectral analysis suggests that the Fe-K emission
is mostly coming from the high ionization layer, which was shocked
earlier (component 4, Table~\ref{multi_pars}). The radial emission profiles
are consistent with this association, but we note that the signal to noise
of the 5-7 keV image on which this is based is low.
The high temperature of this high ionization layer may be surprising, but
it should be mentioned that the electron temperature depends on both
the shock velocity, and the degree of equilibration between electron
and ion temperature. 
In general, one expects that ejecta that is shocked later by the reverse shock
is heated to higher temperatures, but this may be off set by  the shorter
time available for the electrons to equilibrate with the ions.
The situation in 0519-69.0 may somewhat resemble the situation
in the south of Kepler's SNR. The analysis of \citet{chenai04} shows 
that for this region Fe-K peaks at a larger radius than the Fe-L emitting region.

\subsection{Comparison of the stratification with other Type Ia supernova remnants}\label{sec_comparison}
The chemical stratification of SNR~0519-69.0 is striking. Even more so
if one considers that this SNR is much more distant than the well
studied Galactic Type Ia SNRs Tycho (SN 1572), Kepler (SN 1604) and SN 1006.
All these SNRs also show some stratification, but less pronounced.
In some cases the Fe, Si, and O layers seem to be very close
together or even indistinguisable as separate layers.
The latter is especially true for Tycho, in which silicon and
iron are found very close to the shock front \citep{warren05,badenes06}.
For Kepler the situation is different for the northern part of the SNR
than for the southern part \citep{chenai04}. In the north
the spatial separation between Si and Fe is small, whereas in
the south the situation is in fact similar to the stratification
of SNR~0519-69.0.

The situation for SN 1006 is a bit more complicated. There the O-rich
ejecta lie very close to the forward shock \citep{cassamchenai08}, but otherwise
the Fe-rich layer seems to lie at smaller radius, and may not even have been
shocked by the reverse shock, since UV absorption spectra shows the
reverse shock to go through  Si-rich material \citep{hamilton07}.

So what is the origin of this variation in stratification among Type Ia SNRs?
It could be that there is some variation among Type Ia supernovae itself,
with some supernovae producing more mixing than others. 
Optical spectroscopy of Type Ia supernovae suggests a strong radial
stratification \citep[e.g.][]{branch05,mazzali07} with the the outer
ejecta consisting of low mass elements (O, Mg), then followed by 
intermediate mass elements (IME), and the inner ejecta consisting
of Fe and Ni isotopes. However, a group called High Velocity Gradient (HVG) 
supernovae \citep{benetti05} show a strong gradient in the 
Si II velocities as a function of time, and evidence for high velocity
IME. For the HVG SN 2002bo Si appears to be mixed in with O \citep{stehle05}.
The strong stratification observed in SNR~0519-69.0 suggests, therefore, 
that its supernova  did not belong to the  HVG class. 

This does not mean, however, that Kepler or Tycho did belong to the HVG
class, because other mechanisms may have destroyed the strong layering in 
the SNR phase. Kepler is in that sense a very interesting case, as the northern
and southern regions show different behaviors. In the north the blast
wave of Kepler appears to be interacting with a pre-existing shell of 
about 1~M$_\odot$, which caused a strong deceleration of the blast wave 
\citep[][]{vink08b}. The lack of stratification in that
region may therefore be attributed to the strong deceleration, squeezing
the different layers closer together.

Interestingly, the expansion rate of Tycho in the eastern part,
where the ejecta are very close to the shock front, is also much lower
than expected for a SNR expanding in a homogeneous medium \citep{katsuda10}.
It is, therefore, possible that 
also in Tycho the initial stratification is no longer visible,
as the layers have been squeezed together due to 
the presence of a strong density gradient, and perhaps even mixed
due hydrodynamical instabilities.

Another explanation for the proximity of the ejecta
to the shock front in Tycho is the effect of efficient cosmic ray acceleration 
\citep{warren05}. The efficient acceleration alters the equation of state
due to the large pressure fraction attributed to relativistic particles
and energy losses due to escape of the highest energy cosmic rays
\citep{decour00,helder09}. The distance between contact discontinuity 
and shock front is smaller
in the east, whereas the X-ray synchrotron rims, a signature of
cosmic ray acceleration, are more pronounced in the west.
Moreover, one should not confuse the distance {\em between} contact
discontinuity and shock
front, with a lack of stratification {\em within} the ejecta.
Only if the equation of state within the ejecta is changed, can
one explain the lack of stratification by invoking cosmic rays.
This could be the case, if
also the reverse shock is efficiently accelerating particles.
For Tycho there is no evidence that this is the case, but the reverse shock
of the core collapse SNR Cas A has been proposed as a site of efficient
cosmic ray acceleration \citep{helder08}.

In addition to cosmic ray acceleration it has been proposed that 
Rayleigh-Taylor instabilities have brought ejecta close to the shock front
\citep{chevalier92,wang01,badenes06}. If this is indeed the case, it may
also have destroyed the initial stratification of the ejecta. 
It is, however, not clear
why this mechanism would have affected the stratification of Tycho
but not of SNR 0519-69.0, or the southern part of Kepler.
So perhaps it is a combination of factors that made 
Tycho and the north of Kepler less stratified, than 
0519-69.0 and the south of Kepler.

\subsection{The chemical composition of SNR 0519-69.0}
From our spectral analysis we derive ejecta masses for O, Si/S, Ar/Ca, and
Fe. These masses should be regarded with some caution, because inhomogeneities
in the density distribution can lead to some systematic errors. However, 
on face value,  our estimates indicate
that the overall abundance pattern
matches that for normal Type Ia supernovae, but that the oxygen mass
is rather large, $M_{\rm O}\approx 0.2-0.3$~M$_\odot$.

According to a recent analysis of optical spectra of Type Ia supernovae
\citep{mazzali07}
the amount of material from the C/O white dwarf burned into IME and more massive
elements is remarkably homogeneous, corresponding to a total mass of
roughly $M_{\rm IME,Fe,Ni}\approx 1.1$~M$_\odot$.
This means that about 0.2-0.3~M$_\odot$ consists of lower mass elements.
This is consistent with the amount we infer for SNR~0519-69.0, provided
that most of the outer layers consist of oxygen and that little
unburned carbon is present. 

One of the differences
between pure deflagration and
delayed detonation explosion models is that the latter can burn material
all the way to the outer edge of the white dwarf, leaving no
carbon.  Optical spectroscopy of Type Ia supernovae reveal that little
or no carbon is present, with the oxygen abundance a factor 100 to 1000 times
the carbon abundance \citep{marion06}. The relatively high mass
of oxygen in 0519-69.0 is therefore consistent with optical spectroscopy
of Type Ia supernovae, and
supports the idea that burning in the outer layers converts almost all carbon
to oxygen.

The amount of Fe that we infer, $\sim 0.7$~M$_\odot$, is also fairly
typical for Type Ia supernovae. Here we need to caution that not all
the Fe may have been shock heated by the reverse shock, so the total
iron mass may be higher. \citet{mazzali07} showed that the peak
luminosity, for which they use the drop in magnitude within 15 days after maximum ($\Delta m_{15}$)  as a proxy,
correlates with the total amount of iron/nickel synthesized
\citep[see also][]{woosley07}. Using their ``Zorro diagram'' we
can translate our estimate of shocked iron (which includes decayed $^{56}$Ni),
in an estimate for $\Delta m_{15}$. $M_{\rm Fe} >0.7$~M$_\odot$
corresponds to $\Delta m_{15}\lesssim 1.4$.
Combining the amount of IME that we estimate $M_\mt{IME} = 0.2-0.4$~M$_\odot$ with
the ``Zorro-diagram'' suggests that $\Delta m_{15} \gtrsim 1.1$.

These constraints suggest that the supernova was a fairly normal
Type Ia event, not as bright as SN1991T ($\Delta m_{15}= 0.95$), but similar
to for example the template Type Ia 
SN 1996X \citep[$\Delta m_{15}= 1.3$,][]{salvo01}.
This is unlike the youngest known Type Ia SNR in the LMC, 0509-67.5, which was
according to the light echo spectrum of the supernova an exceptionally
bright Type Ia \citep{rest08a}, similar to SN 1991T.

There is some evidence that bright Type Ia supernovae are associated with
younger stellar populations \citep{gallagher08}, but the starforming
history around SNR 0509-67.5 is not significantly younger
than around SNR 0519-69.0 \citep{badenes09}.
This study, therefore, does not provide evidence 
for a one to one
relation between population age and supernova brightness.
In addition, \citet{howell09} show that the relation between the mean galaxy 
stellar population
age and Type Ia brightness has significant scatter. 
This may be no suprise, if 
the brightness of a Type Ia supernova is related to the properties
and history of the progenitor white dwarf, whereas the time delay
between birth of the binary system and the supernova event is also governed
by the evolution of the secondary star \citep{howell09}.\footnote{Nevertheless,
some correlation between time delay and white dwarf properties can be expected,
since a long time scale for the evolution of the primary star (the white
dwarf that will explode), automatically
means that the secondary star evolves on an even longer time scale.}

\subsection{Analytical model}
\begin{figure*} 
   \begin{center}
              \parbox[b]{0.47\hsize}{
                \includegraphics*[width=\hsize,angle=0]{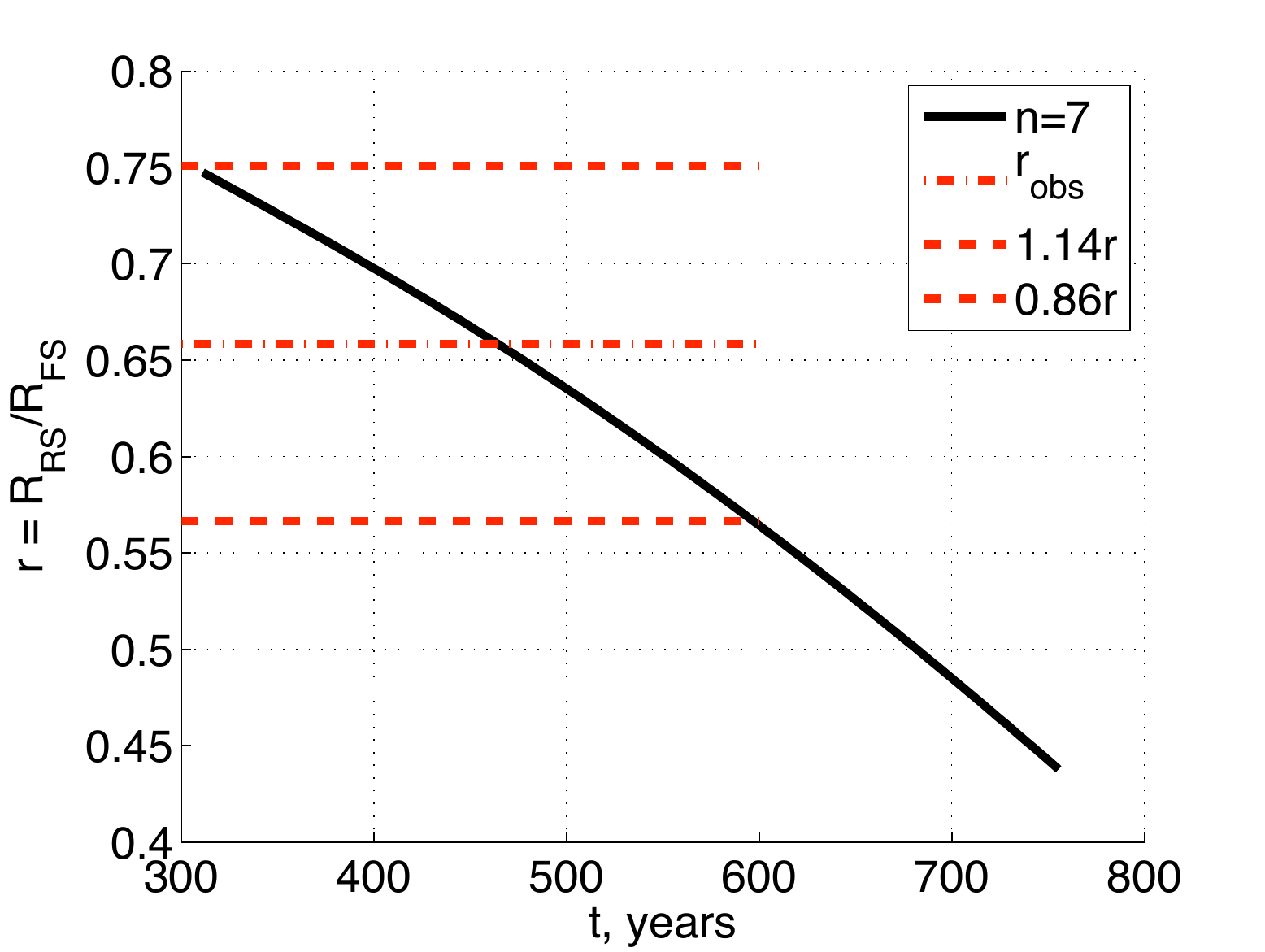} \\
                \includegraphics*[width=\hsize,angle=0]{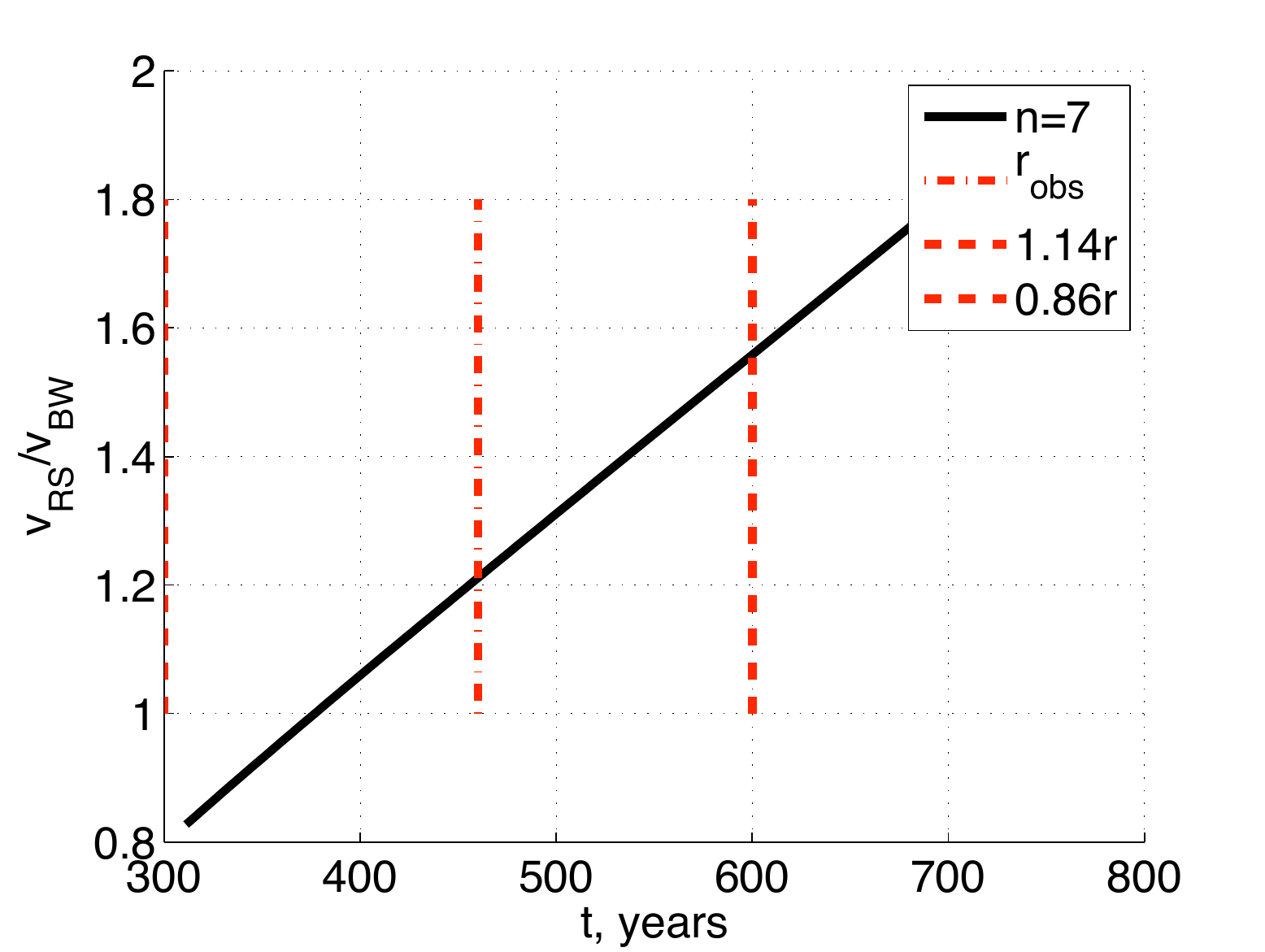}
              }
               \includegraphics*[width=0.52\hsize,angle=0]{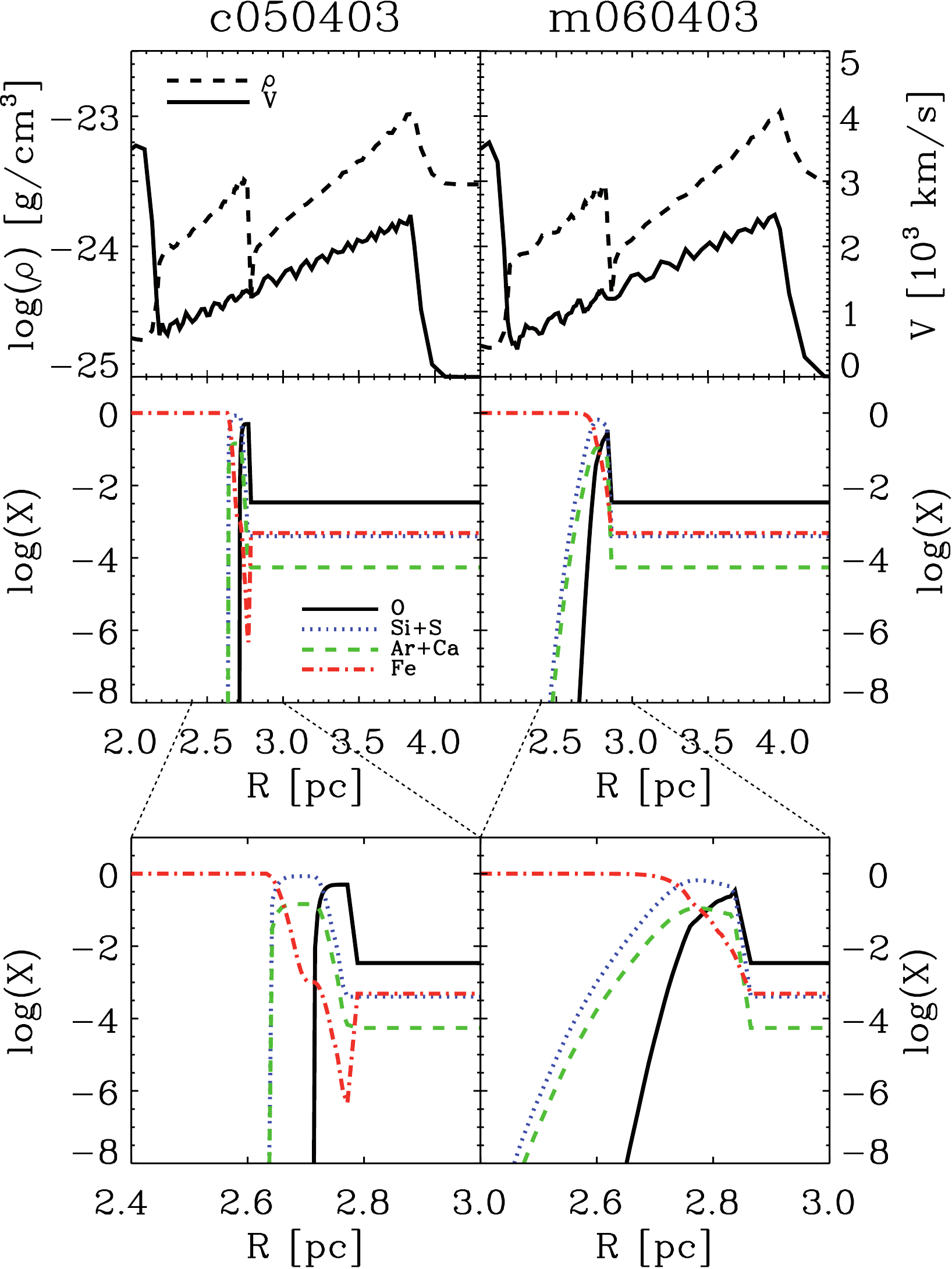}%
    \caption{{\bf Left two plots:} \citet{trmck} analytical solution for SNR~0519-69.0. Radii ratio vs. the age (top panel); red dash-dotted lines show the measured value from the Chandra measurements (Sect.~\ref{sec_profiles}). Reverse shock velocity (in the frame of the ejecta) to forward shock velocity ratio vs. the age (bottom panel); red dash-dotted lines show values of age derived from the top plot. {\bf Right four plots:} density, velocity (top panels) and abundances (bottom panels) profiles of the numerical models. In this set, left plots show the ``mildly mixed'' c050403 model, right plots --- ``moderately mixed'' m060403.}%
    \label{models}%
    \end{center}
  \end{figure*}

From the measured locations of the reverse and forward shocks, we can derive 
typical velocities of the plasma, using the \citet{trmck} analytical models, 
with the ejecta density power $n=7$. Setting the density of the circumstellar 
matter to $n_\mt{CSM} = 2\; \mt{cm}^{-3}$, supernova explosion energy to 
$E = 10^{51}$ erg, and the mass of the progenitor to 
$M_\mt{WD} = 1.4\,M_\odot$, we can plot the forward to reverse shock radii ratio
 versus the time that passed since the explosion. We see from the left panel of
 the \rfig{models} that these parameters correspond to age of the remnant of  
$460\pm150$ years, which is somewhat lower than the value estimated by \citet{rest}. The leftmost bottom panel of 
\rfig{models} shows that in this model, the ratio of the reverse to forward 
shock velocity is $v_{RS}/v_{BW} \simeq 1.2$. 

\subsection{Numerical HD models}
We run several hydrodynamical models \citep{sorokina, kosenko}, in order to reproduce the observed features of SNR 0519-69.0. We set the CSM density to $\rho_\mt{CSM} = 3\times10^{-24}\;\mt{g\,cm}^{-3}$.  We considered two ``toy'' explosion models by \citet{woosley07} of thermonuclear supernovae: a ``mildly mixed'' c050403 ($M_\mt{Ni} = 0.5M_\odot,\;M_\mt{Fe} = 0.4M_\odot,\;M_\mt{IME} = 0.3M_\odot$) and a ``moderately mixed'' m060403  ($M_\mt{Ni} = 0.6M_\odot,\;M_\mt{Fe} = 0.4M_\odot,\;M_\mt{IME} = 0.3M_\odot$). The distribution of the elements in these models is plotted in the right lower panels of \rfig{models}. Velocity and density profiles of the models at the age of 560 years, which are based on these two initial set ups, are presented in \rfig{models} (top graphs).

In the hydrodynamical models the 
contact discontinuity is further away from the blast wave and closer to 
the reverse shock as compared to the observed locations in the SNR 0519-69.0. 
Suggesting that some of the effects  that may have brought the ejecta in
Tycho's SNR close to the forward shock, whether pure hydrodynamical effects,
or efficient cosmic ray acceleration, 
may to a lesser extent also have affected  SNR 0519-69.0.
In the near future we intent to produce models that incorporate
cosmic ray acceleration into the hydrodynamic models (Kosenko et al. in
preparation). 
Note that 
\citet{cassamchenai08} found that for SN\,1006 the contact discontinuity
is also close to the shock front,
to a lesser extent even for those regions that do not
emit X-ray synchrotron radiation, where efficient cosmic ray 
acceleration is likely not important. Their models suggest that
cosmic ray acceleration alone is not sufficient to explain the observations.
This suggests that the distance of the contact
continuity to the shock front is the result of a mixture of efficient cosmic
ray acceleration and hydrodynamical instabilities 
\citep[see also][and the discussion in Sect.~\ref{sec_comparison}]{miceli09}. 

Estimated from our numerical models, the amount of swept up supernova matter is as follows: $M_\mt{O} = 7\%,\; M_\mt{Si/S} = 12\%/7\%,\; M_\mt{Ar/Ca} = 1\%/2\%,\;M_\mt{Fe} = 63\%$ for c050403 and $M_\mt{O} = 3\%,\; M_\mt{Si/S} = 12\%/7\%,\; M_\mt{Ar/Ca} = 1\%/2\%,\;M_\mt{Fe} = 71\%$ for  m060403 explosion model. With a total mass of the shocked ejecta of $1.2\,M_\odot$ in both cases. 
We see that the fractional mass of Fe in c050403 model matches reasonably well the observed fractional
mass as determined in Sect.~\ref{sec_o_si_fe}, but  the observed fractional mass of oxygen appears to be higher than indicated by
the models.  This may hint that the progenitor was an oxygen-rich white dwarf. The lack of neon in the RGS spectra and magnesium in the EPIC spectra still argues in favor of C/O white dwarf, and not an O/Ne/Mg white dwarf, which is consistent with type Ia explosion models \citep[e.g.][]{woosley07}.

\subsection{RGS line broadening and the shock velocity}
Using the measured value of the emission lines broadening ($\sigma_\mt{RGS}$), 
we can estimate the velocity of the forward shock. The temperature of each 
specie behind the reverse shock is calculated from (no temperature 
equilibration between species assumed)
\BE
kT_i = \frac{3}{16}m_i v_\mt{RS}^2
\label{ktvs}
\EE
here $m_i$ --- mass of species $i$, $v_\mt{RS}$ --- reverse shock velocity. Thus,
the Maxwellian thermal broadening of the lines is
\BE
\sigma_\mt{th\, ej} = \sqrt{\frac{kT_i}{m_i}} = \frac{\sqrt{3}}{4}v_\mt{RS}
\label{sigmath}
\EE
As it is shown above, according to the analytical \citet{trmck} solution, for the case of SNR~0519-69.0 the relation $v_\mt{RS} = 1.2\, v_\mt{FS}$ holds. Also, according to the numerical models, the value of the bulk velocity in the ejecta is by  $\sim 50\%$ lower than the bulk velocity in the shocked CSM (the velocity profiles of the HD models on the top right plots of~\rfig{models}), i.e. $\sigma_\mt{bulk\, ej} \simeq 0.5 \sigma_\mt{bulk\, CSM}  = 0.5\,(3/4)\,v_\mt{FS}$. Thus
\BE
\sigma_\mt{RGS} = \sqrt{\sigma_\mt{th\, ej}^2 + \sigma_\mt{bulk\, ej}^2} \simeq 0.64\, v_\mt{FS}
\label{sigmatot}
\EE
Substituting the measured value of $\sigma_\mt{RGS} = 1873\pm 50$~km\,s$^{-1}$\ 
into eq.~(\ref{sigmatot}), we obtain $v_\mt{FS} = 2927\pm 500$~km\,s$^{-1}$. Where the error
includes an estimate of the systematic uncertainties, which come from relation of $\sigma_\mt{RGS}$ to $v_\mt{FS}$.

From the optical observation of the remnant 
we have independent measurements of the blast wave velocity, but these 
measurements give ambiguous values.
\citet{adelsberg08} modeled hydrogen spectra of the SNR and found shock 
velocities of   $v_\mt{FS} = 2984^{+703}_{-185}$  km\,s$^{-1}$\ and 
$v_\mt{FS} = 1178^{+185}_{-157}$   km\,s$^{-1}$\  based on the observations of \citet{ghav07}  
and \citet{smith91} respectively. Measurements of H$\alpha$-line along the rim 
by \citet{tuohy82} provide the value of $v_\mt{FS} = 2800\pm300$  km\,s$^{-1}$. 
Our estimate is the closest to the estimates by \citet{tuohy82} and
\citet{ghav07}. The lower value of the 
forward shock velocity in the measurements by \citet{smith91} is 
probably caused by the slit location they used, which intersects the
bright eastern rim of the SNR, where the blast wave probably 
encounters a dense cloud of interstellar matter and may 
have slowed down \citep{ghav07}.

\section{Conclusions} \label{conclusion}
We analyzed X-ray data of SNR~0519-69.0, which were obtained by both the
XMM-Newton and Chandra observatories. We treated the spectra with {\sc spex} fitting software and applied non-equiliblium ionization (NEI) models. Fitting of the data with the single-ionization timescale NEI model allowed us to measure average electron temperature of $3$ keV, ionization timescale of $2.3\times10^{10}\;\mt{s\;cm}^{-3}$. 

The stratification of the elements is particularly pronounced in 
the radial emissivity profiles of the Chandra images of the SNR. Chandra ACIS
spectra of specific regions of the SNR also points to the separation of the 
elements in the shell. This inspired us to use a multi-component fitting model 
for the EPIC MOS and RGS spectral analysis, in which the observed
layers are each represented by an NEI component dominated by one or two elements.
This multi-component model provides an adequate description of both
the broad band EPIC MOS spectra  and the high spectral resolution 
RGS spectra in the 0.5-1.1 keV band. In addition it allowed us
 to characterize the Chandra ACIS spectra of several distinct regions.
 Using this approach we estimate the density of the circumstellar matter of $2.4\pm0.2\;\mt{cm}^{-3}$. 

The results of our analysis show that the X-ray data of SNR~0519-69.0 are 
consistent with 
mildly or moderately mixed delayed detonation explosion models, which result 
in clear 
separation of the outermost oxygen layer (produced during the 
final detonation phase), followed inwards by the partially mixed inner layers 
of intermediate mass elements and iron (deflagration phase of the explosion). 
The mass fraction that we infer are 
$M_\mt{O} \approx 32$\%, $M_\mt{Si/S} \approx 7\%/5\%$,
$M_\mt{Ar+Ca} \approx 1$\% and $M_\mt{Fe} \approx 55$\%, which should
be considered with some caution, as we had to infer a volume filling factor
to convert from emission measure to mass. 
Taken at face value, these mass fraction are consistent with
recent observational results of Type Ia supernovae \citep{mazzali07}, 
which show that
all material out to an enclosed mass of $\sim 1.1$~M$_\odot$ is burned into
IME and heavier elements, leaving about $0.2-0.3$~M$_\odot$ for oxygen,
provided all carbon was burned.
The mass fractions of IME and Fe in SNR 0519-69.0 puts contraints on
the sub-classification of the supernova, suggesting a Type Ia
event of normal peak luminosity.

The RGS data allowed us to measure the line broadening of the shocked ejecta of
 $1873\pm 50$ km\,s$^{-3}$, which helped to constrain the dynamical properties 
of the remnant and estimate the forward shock velocity of $2770\pm 500$ km\,s$^{-1}$. 
This value is consistent with the measurements reported by \citet{tuohy82} and \citet{ghav07}, 
but inconsistent with \citet{smith91}.

\begin{acknowledgements}
We thank the anonymous referee for his/her insightful comments, which helped
to broaden the scope of this paper.
We also thank S.I.~Blinnikov for providing us with the thermonuclear 
explosion models and  also for the fruitful discussions on explosion mechanisms.
DK is supported by an ``open competition'' grant and EH and JV are supported 
by a Vidi grant  from the Netherlands Organization for Scientific Research 
(PI J.~Vink). The work is also partially supported 
in Russia by RFBR under grant  07-02-00961.
\end{acknowledgements}

\bibliographystyle{aa}
\bibliography{main}

\end{document}